\documentclass[sigconf,authorversion,nonacm]{acmart}
\usepackage{balance}
\usepackage[utf8]{inputenc}%(only for the pdftex engine)
\def\BibTeX{{\rm B\kern-.05em{\sc i\kern-.025em b}\kern-.08emT\kern-.1667em\lower.7ex\hbox{E}\kern-.125emX}}

\usepackage{longtable}
\usepackage{tabularx}
\usepackage{afterpage}
\usepackage{placeins}
\usepackage{rotating}
\usepackage{array}
\usepackage{todonotes}
\usepackage{xspace}
\usepackage[linesnumbered,ruled,vlined]{algorithm2e}
\usepackage{adjustbox}
\usepackage[english]{babel}
\usepackage{soul}
\usepackage{flushend}
\usepackage{listings}
\usepackage{amsfonts}
\usepackage{graphicx}
\usepackage{subcaption}
\usepackage{multirow}
\usepackage{latexsym}
\usepackage{fmtcount}
\usepackage{courier}

\theoremstyle{definition}

\usepackage{amsmath}

\usepackage{hyperref}
\usepackage{url}

\usepackage{breakurl}
\usepackage{cleveref}
\Crefformat{figure}{Fig.~#2#1#3}
\Crefmultiformat{figure}{Figs.~#2#1#3}{ and~#2#1#3}{, #2#1#3}{ and~#2#1#3}

\makeatletter
\patchcmd{\@makecaption}
  {\scshape}
  {}
  {}
  {}

\newcommand*{\da@rightarrow}{\mathchar"0\hexnumber@\symAMSa 4B }
\newcommand*{\da@leftarrow}{\mathchar"0\hexnumber@\symAMSa 4C }
\newcommand*{\xdashrightarrow}[2][]{%
  \mathrel{%
    \mathpalette{\da@xarrow{#1}{#2}{}\da@rightarrow{\,}{}}{}%
  }%
}
\newcommand{\xdashleftarrow}[2][]{%
  \mathrel{%
    \mathpalette{\da@xarrow{#1}{#2}\da@leftarrow{}{}{\,}}{}%
  }%
}
\newcommand*{\da@xarrow}[7]{%
  \sbox0{$\ifx#7\scriptstyle\scriptscriptstyle\else\scriptstyle\fi#5#1#6\m@th$}%
  \sbox2{$\ifx#7\scriptstyle\scriptscriptstyle\else\scriptstyle\fi#5#2#6\m@th$}%
  \sbox4{$#7\dabar@\m@th$}%
  \dimen@=\wd0 %
  \ifdim\wd2 >\dimen@
    \dimen@=\wd2 %   
  \fi
  \count@=2 %
  \def\da@bars{\dabar@\dabar@}%
  \@whiledim\count@\wd4<\dimen@\do{%
    \advance\count@\@ne
    \expandafter\def\expandafter\da@bars\expandafter{%
      \da@bars
      \dabar@ 
    }%
  }%  
  \mathrel{#3}%
  \mathrel{%   
    \mathop{\da@bars}\limits
    \ifx\\#1\\%
    \else
      _{\copy0}%
    \fi
    \ifx\\#2\\%
    \else
      ^{\copy2}%
    \fi
  }%   
  \mathrel{#4}%
}
\makeatother

\newcolumntype{M}[1]{>{\centering\arraybackslash}p{#1}}

\usepackage{enumitem}
\setlist[itemize,1]{leftmargin=1.5\parindent, itemsep=0ex, topsep=0.5ex, % bottomsep=0.5ex
}
\setlist[enumerate,1]{leftmargin=2\parindent, itemsep=0ex, topsep=0.5ex, % bottomsep=0.5ex
}

\setcopyright{none} % switch off copyright text
\begin{document}

%%
%% The "title" command has an optional parameter,
%% allowing the author to define a "short title" to be used in page headers.
\title{{\bf DP-UTIL}: Comprehensive Utility Analysis of\\ Differential Privacy in Machine Learning}

\author{Ismat Jarin}
\affiliation{%
 \institution{University of Michigan, Dearborn}
 \country{United States}}
 \email{ijarin@umich.edu}

\author{Birhanu Eshete}
\affiliation{%
  \institution{University of Michigan, Dearborn}
  \country{United States}}
  \email{birhanu@umich.edu}

\begin{abstract}
Differential Privacy (DP) has emerged as a rigorous formalism to reason about quantifiable privacy leakage. In machine learning (ML), DP has been employed to limit inference/disclosure of training examples. Prior work leveraged DP across the ML pipeline, albeit in isolation, often focusing on mechanisms such as gradient perturbation.

In this paper, we present, DP-UTIL, a holistic utility analysis framework of DP across the ML pipeline with focus on input perturbation, objective perturbation, gradient perturbation, output perturbation, and prediction perturbation. Given an ML task on  privacy-sensitive data, DP-UTIL enables a ML privacy practitioner perform holistic comparative analysis on the impact of DP in these five perturbation spots, measured in terms of model utility loss, privacy leakage, and the number of truly revealed training samples.

We evaluate DP-UTIL over classification tasks on vision, medical, and financial datasets, using two representative learning algorithms (logistic regression and deep neural network) against membership inference attack as a case study attack. One of the highlights of our results is that prediction perturbation consistently achieves the lowest utility loss on all models across all datasets. In logistic regression models, objective  perturbation results in lowest privacy  leakage  compared  to  other perturbation techniques. For deep neural networks, gradient perturbation results in lowest privacy leakage. Moreover, our results on true revealed records suggest that as privacy leakage increases a differentially private model reveals more number of member samples. Overall, our findings suggest that to make informed decisions as to which perturbation mechanism to use, a ML privacy practitioner needs to examine the dynamics between optimization techniques (convex vs. non-convex), perturbation mechanisms, number of classes, and privacy budget.
\end{abstract}

\settopmatter{printfolios=true}
\maketitle

\section{Introduction}
\label{sec: intro}

The continued progress in machine learning (ML) has resulted in a line of applications where privacy-sensitive datasets (e.g., medical records) are increasingly being used to train and deploy ML models. Almost parallel to the ML progress, specially with the impressive performance of deep learning models, privacy attacks against ML models also emerged. Prior work has shown that ML models trained on privacy-sensitive data are vulnerable to a range of privacy-motivated attacks such as membership inference~\cite{MIA_shokri}, attribute inference~\cite{Property_Inference1}, model inversion~\cite{model-inversion}, and model parameters inference ~\cite{model-stealing16}. Moreover, unintended memorization of privacy-sensitive details at training time results in inadvertent leakage of private data at prediction time~\cite{Unintentional-Memorization1,Unintended_Memorization}. A practical challenge in the context of ML is that of striking a reasonable balance between the utility (e.g., accuracy) of the ML model and the privacy of subjects from whom training data is obtained.

To counter ML privacy attacks, differential privacy (DP)~\cite{DP-Dwork} has emerged as a rigorous notion to formalize and measure privacy guarantee based on a parameter called privacy budget.
Across the ML pipeline, DP has been used to limit inference/disclosure of ML training examples pre-training (via input perturbation~\cite{Input-Perturb13,input-perturb20}), during training (via objective perturbation~\cite{ERM-DP} and gradient perturbation~\cite{DP-SGD}), and post-training (via output perturbation~\cite{ERM-DP} and prediction perturbation~\cite{PRICURE21,PATE17,PATE_new}).  

\textbf{Motivation:} While prior work~\cite{ERM-DP,Input-Perturb13,PRICURE21,DP_USENIX,input-perturb20,PATE17,PATE_new} has leveraged these perturbation mechanisms across the ML pipeline, these applications were done in isolation, often focused on one of the perturbation methods (e.g., gradient perturbation). As a result, given a privacy-sensitive dataset and a ML task (e.g, medical image classification), there is lack of a holistic assessment methodology as to the utility of DP when it is employed before, during, and after training. More precisely, isolated applications of DP in prior work do not shed light on how the aforementioned perturbation methods compare in their effectiveness and, more importantly, in their trade-offs. Moreover, the usefulness of these alternative perturbation methods across diverse datasets and ML models remains under-explored.

\textbf{DP-UTIL Overview:} In this paper, we present DP-Util --a framework for holistic utility analysis of DP across the ML pipeline. DP-UTIL enables a ML privacy practitioner to analyze perturbation methods in terms of their impact on model utility, privacy leakage, and actual number of privacy-sensitive samples inferred by an adversary. The benefit DP-UTIL is twofold. First, it enables a ML privacy practitioner to have an {\em across-the-ML-pipeline view} of the impact of DP using standard metrics such as model accuracy and privacy guarantee/leakage. Second, it enables {\em comparative analysis} on the suitability of one DP application (e.g., objective perturbation) against another DP application (e.g., gradient perturbation) so that the practitioner makes informed decisions as to where to plug DP in the ML pipeline.

\begin{table*}[t!]

\scalebox{0.69}{
  \begin{tabular}{lcccccccccccccc}
    \toprule
    \multirow{2}{*}{\bf Work} &
      \multicolumn{5}{c|}{\bf DP Perturbation Type} &
      \multicolumn{3}{c|}{\bf Model} &
      \multicolumn{2}{c|}{\bf Case Study Attack} & 
    \multicolumn{4}{c}{\bf Metric}\\\cline{2-15}
    
      & Input & Objective & Gradient & Output & Prediction & NB & LR & DNN & Membership Inference & Attribute Inference & Utility  Loss & Privacy Leakage & True Positive & Privacy Budget \\
      \midrule
      
    \cite{DP_USENIX} & -- & -- & $\checkmark$ & -- & -- & -- & $\checkmark$ & $\checkmark$ & $\checkmark$ & $\checkmark$ & $\checkmark$ & $\checkmark$ & $\checkmark$ & $[10^{-2},10^{3}]$\\
    
    \cite{input-perturb20} & $\checkmark$ & -- & -- & -- & -- & -- & $\checkmark$ & $\checkmark$ & -- & -- & $\checkmark$ & -- & -- & $[10^{-2},0.25]$\\
    
    \cite{DPUtility20} & $\checkmark$ & -- & $\checkmark$ & $\checkmark$ & -- & $\checkmark$ & -- & $\checkmark$ & $\checkmark$ & $\checkmark$ & $\checkmark$ & $\checkmark$ & -- & $[10^{-2},10^{3}]$\\
    
    DP-UTIL & $\checkmark$ & $\checkmark$ & $\checkmark$ & $\checkmark$ & $\checkmark$ & -- & $\checkmark$ & $\checkmark$ & $\checkmark$ & -- & $\checkmark$ & $\checkmark$ & $\checkmark$ & $[10^{-2},10^{4}]$\\
    
    \bottomrule
  \vspace{1em}
  \end{tabular}}
\caption{Comparison with closely related work. \textbf{NB}: Naive Bayes. \textbf{LR}: Logistic Regression.  \textbf{DNN}: Deep Neural Network.}
    \label{tab:comparison}
\end{table*}

We evaluate DP-UTIL over classification tasks on three datasets covering vision, medical, and financial domains. We use membership inference as a case study attack to analyze privacy leakage and actual number of revealed data samples. To shed light on the difference between convex and non-convex optimization formulations used in training ML models, we use Logistic Regression (LR) and Deep Neural Network (DNN), respectively, as representative ML models due to the wide usage of both on privacy-sensitive datasets. For LR models, we evaluate five perturbation methods: input perturbation~\cite{Input_perturb17}, objective perturbation~\cite{ERM-DP}, gradient perturbation~\cite{DP-SGD}, output perturbation~\cite{ERM-DP}, and prediction perturbation~\cite{PATE17,PATE_new}. For DNN models, we compare input perturbation~\cite{input-perturb20}, gradient perturbation, and prediction perturbation because these three are widely implemented for DNNs.
 
\textbf{Main Findings:} Our findings suggest that perturbation techniques that offer lower utility loss are more vulnerable to inference attack. Moreover, for lower privacy budget, perturbation techniques like objective perturbation and output perturbation result in ML models that classify near random guessing, i.e., produce extreme utility loss that models fail to classify correctly. For binary classifiers, objective perturbation is a better choice compared to gradient perturbation while for multi-class classifiers, objective perturbation offers the highest privacy/utility trade-off. For multi-class classifiers, gradient perturbation performs well in terms of privacy/utility trade-off. Over all model architectures and datasets, prediction perturbation results in lowest utility loss but at a cost of privacy leakage. True revealed records has almost a linear relationship with privacy leakage. Over all the results, we observe that as the privacy leakage increases, a model starts to leak more true records. In a nutshell, our detailed evaluations suggest that, to make informed decisions as to which perturbation mechanism to use, a ML privacy practitioner needs to examine the dynamics between optimization techniques (e.g., convex vs. non-convex), perturbation mechanisms, number of classes (e.g., binary vs. multi-class), and privacy budget.

\textbf{Comparison with Closely Related Work:} DP-UTIL complements prior work in two major ways. First it enables {\em comprehensive DP utility analysis} covering five DP perturbation mechanisms for LR and three DP perturbation methods for DNN. Second, it {\em sheds new light} on the utility of DP in the ML pipeline.\\
\textbf{\em DP-UTIL is more comprehensive}: Table \ref{tab:comparison} summarizes the coverage comparison of DP-UTIL and closely related work\cite{DP_USENIX,DPUtility20,input-perturb20}. 
Compared to~\cite{DP_USENIX} which is limited to utility analysis of gradient perturbation for DP, DP-UTIL covers all the 5 perturbation methods for LR and 3 widely used perturbations for DNN. Hence, it is more comprehensive. In addition, while ~\cite{DP_USENIX} uses image classification datasets, we evaluate DP-UTIL with two more datasets from medical and finance domains in addition to a benchmark image classification dataset.\\ 
With respect to~\cite{input-perturb20} which studies the privacy guarantee offered by input perturbation against objective, gradient, and output perturbation, in DP-UTIL we extend the analysis with prediction perturbation and also extend the evaluation metrics with privacy leakage and number of truly revealed training examples over a wider range of privacy budget than ~\cite{input-perturb20}. Additionally, ~\cite{input-perturb20} does not offer deeper insights on implications of considering different experimental setups (e.g., binary vs. multi-class models, LR vs. DNN, image data vs. numerical data).\\
Compared to~\cite{DPUtility20} which covers input and gradient perturbation for DNN and input and output perturbation for Naive-Bayes, DP-UTIL extends the coverage by analyzing three more DP perturbations for LR and one more (prediction perturbation) for DNN. 

\textbf{\em DP-UTIL offers new insights}:
In ~\cite{DP_USENIX}, the main takeaway is that relaxed DP formulations improve model utility for a given privacy budget, yet the lower DP noise results in additional privacy leakage (hence, the utility does not come for free). Compared to ~\cite{DP_USENIX}, we observe in some cases, a perturbation method results in lower utility over other perturbation techniques, hence costs privacy leakage in exchange. Thus, choosing one perturbation method over another for better utility does not come without paying in privacy leakage. For example, prediction perturbation offers the lowest utility loss for both LR and DNN, hence costs more privacy leakage compared to other perturbation mechanisms.\\
%In ~\cite{DP_USENIX}, the main take-away is that relaxed DP formulations improve model utility for a given privacy budget, but the lower DP noise results in additional privacy leakage (hence, utility does not come for free). Compared to ~\cite{DP_USENIX}, we observe a perturbation method can offer lower utility compared to other perturbation techniques for roughly the same privacy budget. For example, our findings suggest that prediction perturbation offers the lowest utility loss for both LR and DNN compared to other perturbation mechanisms.\\
Compared to ~\cite{input-perturb20}, where theoretical guarantee for input perturbation seems promising, our experimental findings suggest that input perturbation results in rapid privacy leakage with higher privacy budget and this change is usually triggered at $\epsilon \geq 1$.\\
In ~\cite{DPUtility20}, their findings suggest that the number of classes of a given dataset is unlikely to influence where the privacy/utility trade-off occurs. Our findings rather suggest that number of classes has implications on privacy/utility trade-off. For objective perturbation, for instance, binary classifiers show overall better privacy/utility trade-off compared to multi-class classifiers for both DNN and LR models. Our evaluations also suggest overall similar findings for gradient and input perturbation. Another major conclusion in ~\cite{DPUtility20} is noise added at a later stage (e.g., output) in the ML pipeline results in lower utility loss. However, our findings show that objective/gradient perturbation overall results in lower utility loss compared to output perturbation.

In summary, this paper makes the following contributions:
\begin{itemize}
    \item We propose, DP-UTIL, a holistic utility analysis framework for differential privacy across the machine learning pipeline to understand the impact of different perturbation techniques with respect a given range of privacy budget (Sections \ref{subsec:utility-analysis}, \ref{subsec:privacy-leakage-analysis}, and \ref{subsec:true-positive-analysis}). To that end, we analyze input perturbation, objective perturbation, gradient perturbation, output perturbation, and prediction perturbation.
    
    \item Using membership inference as a case study and privacy leakage as a metric, we comparatively analyze the extent to which machine learning models are protected with  state-of-art DP perturbation techniques (Section \ref{subsec:privacy-leakage-analysis}). 
    
    \item We perform a comprehensive study of utility loss and privacy leakage over a range of privacy budget values for two model architectures (Logistic Regression and Deep Neural Network), two naturally privacy-sensitive datasets (finance: LendingClub-Loan dataset~\cite{LendingClub}, healthcare: COVID-19 dataset~\cite{Covid-19}), and a benchmark image classification dataset (CIFAR-10 dataset~\cite{Cifar10}). 
    
    \item We make available our code and data with directions to repeat our experiments. Our artifacts are available for download at: {\color{blue} \url{https://github.com/um-dsp/DP-UTIL}}.
\end{itemize}

The rest of the paper is organized as follows. Section \ref{sec: bground} introduces ML and DP background. Section \ref{sec:perturbations} presents an overview of DP perturbation mechanisms. In Section \ref{sec: design}, we present an overview of the DP-UTIL framework. Our datasets and setup are presented in Section ~\ref{sec: dataset}. Section \ref{sec: eval} presents our findings focusing on utility loss, privacy leakage, and true revealed records. Section \ref{sec: related} surveys closely related work and Section \ref{sec: concl} concludes the paper.
\section{Background}
\label{sec: bground}
In this section, we briefly highlight machine learning preliminaries and the definition of differential privacy.

\subsection{Machine Learning Preliminaries}

 \textbf{Typical ML Training.} In this paper, we focus on supervised machine learning models. Given a set of labeled training samples $X_{train} = (X_i,y_i): i\le n$, where $X_i$ is a training example and $y_i$ is the corresponding label, the objective of training a ML model $\theta$ is to minimize the expected loss over all $(X_i,y_i): J(\theta) = \frac{1}{n}\sum_{1}^{n}l(\theta,X_i,y_i)$. In ML models such as logistic regression and deep neural networks, the loss minimization problem is typically solved using stochastic gradient descent (SGD) by iteratively updating $\theta$ as:
 \begin{equation}\label{eq:gradient-descent}
     \theta = \theta - \epsilon\cdot \Delta_{\theta}\sum_{i=1}^{n} l(\theta,X_i,y_i)
 \end{equation}
 where $\Delta_{\theta}$ is the gradient of the loss with respect to the weights $\theta$; $X$ is a randomly selected set (e.g., {\em mini-batches}) of training examples drawn from $X_{train}$; and $\epsilon$ is the {\em learning rate} which controls the magnitude of change on $\theta$.

 \textbf{Typical ML Testing:} Let $X$ be a $d$-dimensional feature space and $Y$ be a $k$-dimensional output space, with underlying probability distribution $Pr(X,Y)$, where $X$ and $Y$ are random variables for the feature vectors and the classes (labels) of data, respectively. The objective of testing a ML model is to perform the mapping  $f_{\theta} : X \rightarrow Y$. The output of $f_{\theta}$ is a $k$-dimensional vector and each dimension represents the probability of input belonging to the corresponding class. 

\subsection{Differential Privacy}
For two neighboring datasets $D_1$ and $D_1$ which differ by just one data point, let the output space of a randomized mechanism $M$ be $S$. Differential privacy (DP) guarantees that a randomized mechanism $M$ does not enable an observer (adversary) to distinguish whether $M$'s output was based on $D_1$ or $D_2$. Dwork et al.~\cite{Advanced-Comp} formalize $(\epsilon,\delta)$-DP as follows. A mechanism $M$ preserves $(\epsilon,\delta)$-DP if:
 \begin{equation} \label{eq:dp}
Pr[M(D_1) \in S] \le e^\epsilon \times Pr[M(D_2) \in S]+ \delta
\end{equation}
where $\epsilon$ is the privacy budget and $\delta$ is the mechanism's failure probability. When $\delta = 0$, we obtain a strict $\epsilon$-DP formulation of (\ref{eq:dp}). The lower the value of $\epsilon$, the stronger the privacy protection and the higher the utility loss. 

To achieve $\epsilon$-DP, Laplace distribution is a common choice to sample noise. For $(\epsilon,\delta)$-DP, Gaussian distribution allows sampling noise. In both $\epsilon$-DP and $(\epsilon,\delta)$-DP, the sampled noise is correlated with the sensitivity of the mechanism $M$. For two neighboring datasets $D_1$ and $D_2$ differing by one record, the sensitivity $\Delta_{M}$ is the maximum change in the output of M over all possible inputs. Computing $\Delta_{M}$ as the maximum of $||M(D_1)-M(D_2)||$ establishes worst-case upper bound on how much the output of $M$ changes when $D_1$ and $D_2$ are identical except for one record, i.e., $||D_1-D_2||_1=1$.

\section{Perturbation Mechanisms}
\label{sec:perturbations}
In this section, we introduce the five privacy noise mechanisms across the ML pipeline. To guide the forthcoming discussion, we use Algorithm 1 as a high-level skeleton for candidate spots as to where to add DP perturbations. As noted in~\cite{DP_USENIX}, the type (convex or non-convex) of the optimization problem dictates the specifics of DP perturbation mechanisms.

\begin{centering}
\fbox{
\parbox{0.9\columnwidth}{
\textbf{Training Data:$(X_{train},y_{train})$}\\
\textbf{Input perturbation:}\\
$X_i = X_i +noise, \forall X_i \in X_{train}$\\
$\theta=0$\\
\textbf{Objective Perturbation:}\\
$J(\theta)$ =$\frac{1}{n}$ ${\sum{(l(\theta,X_i,y_i)}+ \lambda R(\theta)+noise }$\\
for each epoch:\\
  \phantom{x}\hspace{3ex}\textbf{Gradient Perturbation:}\\
  \phantom{x}\hspace{3ex}$\theta=\theta -\iota(\nabla J(\theta)+noise)$\\
\textbf{Output Perturbation:}\\
$\theta=\theta+noise$ \\
\textbf{Prediction Perturbation:}\\
$y=f_{\theta}(X_i,\theta)+noise$
}
}
\\ \indent \textbf{Algorithm 1:} Perturbations across the ML pipeline.
\end{centering}

\subsection{Input Perturbation}
In a pre-training setting, one natural perturbation alternative is to add noise to individual training examples and produce the perturbed version of $X_{train}$ and train the model on it~\cite{Input-Perturb13,Input_perturb17,input-perturb20}. For a training data $X_{train}$ with dimension $d$, a typical input perturbation on sample $X_i$ is done as:
\begin{equation}\label{eq:input_pert}
X'_i = X_i +noise, \forall X_i \in X_{train}
\end{equation} 
where $noise = X_i^j +Lap(\frac{S_j}{\epsilon/d})$ with $S_j$ as the sensitivity (value range) of the $j^{th}$ feature of $X_i$. We note that to keep the perturbed features within valid value boundaries, clipping is applied using the upper- and lower-bounds of each feature value. In addition, no or weak inter-feature dependency is assumed for such a perturbation to be useful.

% \textbf{To Clarify: Is the above formulation of noise dependent on whether the features are independent from each other? Depending on the absence or presence of dependency between features, the amount of noise may need to be changed. For instance, if features are independent, then adding the maximum possible noise will maximize the privacy guarantee provided that it doesn't result in poor utility.}

When features have diverse representations and unbounded value ranges, estimating sensitivity is not trivial. Domains such as image classification, where features are homogeneously constituted (e.g., pixel intensity values), estimating sensitivity is relatively facile. Another challenge with input perturbation is that post-perturbation, the utility of the trained model needs to be within acceptable utility loss penalty. Given the feature-level fidelity of input perturbation, achieving an acceptable trade-off on model utility is an optimization challenge. Input perturbation has been recently shown \cite{input-perturb20} \cite{Input_perturb17} to offer both local and model privacy guarantees compared with the other perturbation mechanisms. 

Differentially private ERM with input perturbation ensures both local and model privacy. In \cite{Input_perturb17}, it is shown that adding noise to input data depends on privacy parameters ($\epsilon, \delta$), data size $n$, and constants of loss function. They have also shown that this technique satisfies ($\alpha \epsilon, \delta)$-DP where $\alpha$ is learning rate. For Equation \ref{eq:input_pert}, Gaussian noise can be expressed as $\textit{N}(0, \frac{\sigma^2}{n})$,  where $\sigma$ is $\frac{\sqrt{2d}a\lambda+\sqrt{2da^2\lambda^2+(2\lambda(1-2a)/\epsilon}}{(1-2a)}$, $a=\sqrt{\frac{4/\delta}{n}}$, and $d$ is data dimension.

For deep learning, performing input perturbation assumes that the loss function is not  strongly convex though it is G-Lipschitz and satisfies Polyak-Lojasiewicz condition. For Equation \ref{eq:input_pert}, Gaussian noise can be expressed as $\textit{N}(0, \sigma^2)$, where $\sigma^2$ is $c\frac{G^2Tlog(\frac{1}{\delta})}{n(n-1)\epsilon^2}$ for some constant $c$ with $T$ as the total number of iterations \cite{input-perturb20}.

\subsection{Objective Perturbation}
During the training of an ML model, one of the DP perturbation alternatives is {\em objective perturbation}, which works well with convex optimization problems such as the empirical risk minimization (ERM) algorithm~\cite{ERM-DP}. Their technique of objective perturbation is a two-stage process: to add noise to the objective function itself and then revealing the minima of the perturbed objective. For convex optimization problems, suppose we consider logistic regression with $l_2$ regularization penalty. The (convex)objective function $l(.)$ with objective perturbation is computed as:
\begin{equation}\label{eq:loss}
    J(\theta) =\frac{1}{n} \sum_{i=1}^{n} l(\theta,X_i,y_i)+ \lambda R(\theta) + noise
\end{equation}
where $R(.)$ is the regularization function such as $l_1$ and $l_2$.
To add differential privacy guarantee to the model using objective perturbation, a noise is added to $J(\theta)$ and then $\theta =min_{\theta}J(\theta)$ is computed via iterative gradient update using (\ref{eq:gradient-descent}). Chaudhuri et al.~\cite{ERM-DP} prove that if $||X_i||_2\leq 1$ and $y_i \in \{-1,1\}$ then $noise = \frac{2}{n\epsilon}$ is added to the objective function which has a sensitivity of $\frac{2}{n}$.

\subsection{Gradient Perturbation}
The other commonly used perturbation mechanism during training is {\em gradient perturbation}. Again, considering logistic regression with $l_2$ regularization penalty, the gradient of the objective function with gradient perturbation is computed as: 
\begin{equation}
    \nabla J(\theta)=\frac{1}{n}\sum_{i=1}^{n} \frac{-X_i y_i}{1+\exp^{X_i\theta y_i}} +\lambda\theta +noise
\end{equation}
where the gradient has a sensitivity of $\frac{2}{n}$. In gradient descent, the gradient value is computed for each iteration of the training process, which requires sampling noise with a scale of $\frac{2}{n\epsilon}$ for each iteration of model update~\cite{Private_logistic_rg}.
 Now integrated into Google's TensorFlow framework, Abadi et al.~\cite{DP-SGD} proposed DP-SGD, a deep learning training algorithm widely adopted for gradient perturbation of non-convex optimizers. They have used the gradient clipping technique to limit the sensitivity of the training algorithm. Two modifications have been made to ensure that SGD is a differentially private algorithm. First, they bounded the sensitivity of the gradient by clipping them. Second, they have sampled random noise and add to the clipped gradient. Given training data $(X_1, X_2,...,X_n)$ and target labels $(y_1,y_2,...,y_n)$, and gradient $g(X_i)$, to build a ($\epsilon,\delta$)-DP model using SGD technique, this method computes the gradient for a random subset of examples for each batch lot $L$ , clip the $l_2$ norm for each gradient, and add random noise of the distribution of $\it{N}(0,\sigma^2C^2)$, where $C$ is the clipping threshold and $\sigma$ can be expressed as:
 \begin{equation}
    \sigma \geq c_2\frac{q\sqrt{T\log(1/\delta)}}{\epsilon}
\end{equation}
 where $q=\frac{L}{n}$ , $T$ is the step size, and $c_2$ is existing constant. 
 
 %In DP-SGD \cite{DP-SGD}, the privacy amplification theorem implies that each step is $(O(q\epsilon),q\delta)$-DP with respect to the whole dataset, where $q=\frac{L}{n}$ ($n$: size of the input dataset, $L$: lot/batch size) and $\epsilon\le1$. If the number of epochs $E$ and total number of training steps is $T$, where $T=E/q$, an algorithm is said to be $(\epsilon,\delta)$-DP if we choose, 

\subsection{Output Perturbation}
In a post-training setting, output perturbation is used to limit the leakage/inference of true model parameters.  As shown in~\cite{ERM-DP} for convex optimization problem (e.g., ERM), noise is added to the model parameters $\theta$ as follows: 
\begin{equation}
    \theta=\theta + noise
\end{equation}
For logistic regression with $l_2$ regularization, output perturbation typically requires a sampling noise of $\frac{2} {n\lambda \epsilon}$.

\subsection{Prediction Perturbation}
The other alternative perturbation mechanism in a post-training setting is {\em prediction perturbation} whereby random noise is added to the prediction result before producing the final label. For instance, in MemGuard~\cite{Memguard}, random noise is added to the confidence vector to mask the prediction confidence against membership inference attacks of the likes of Shokri et al.~\cite{MIA_shokri}. In PATE~\cite{PATE17, PATE_new}, noise is appended to the majority vote count of model prediction results of teacher ensembles.  In PATE, the training dataset $D$ has been split into $M$ number of disjoint datasets. With these disjoint datasets, $M$ number of models are trained named as teacher models. For an input sample $x$, the final output is chosen via a noisy vote aggregation of the teachers' prediction results as follows: 
\begin{equation}
    f(x)=\operatorname*{argmax}_{j} {v_j(x)+noise}
\end{equation}
where $v_j(x)$ is the number of teachers that assigned class $j\in[1, ..., c]$ to input sample $x$ out of the possible $c$ labels. Given that noise is added on top of a vote count, the prediction is perturbed by sampling noise from Laplace distribution as $Lap(\frac{1}{\epsilon})$ with sensitivity = 1.

\section{DP-Util Design}
\label{sec: design}
\begin{figure*}[t!]
    \centering
    \includegraphics[width=\textwidth]{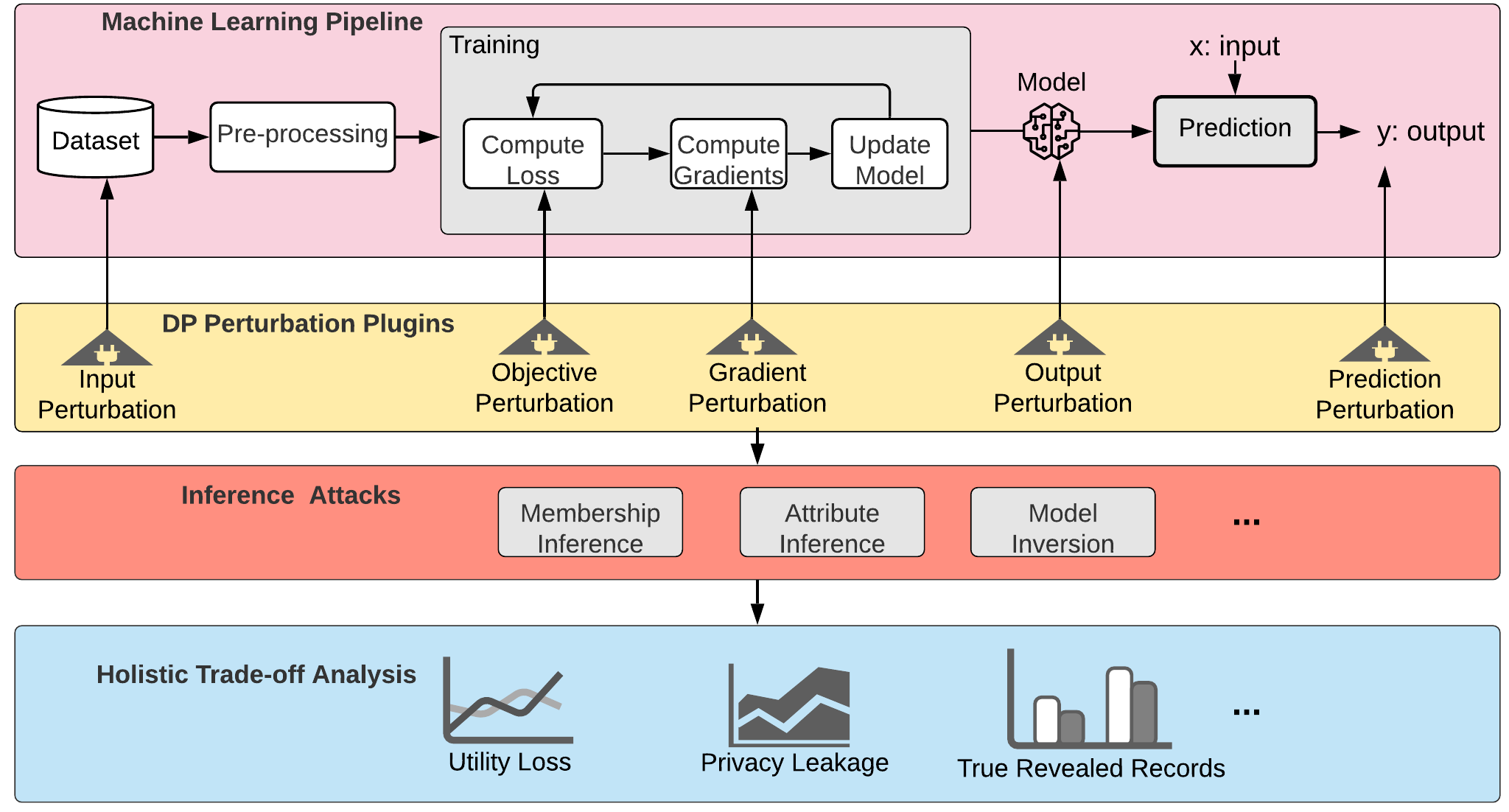}
     \vspace{1em}
    \caption{Overview of DP-Util Pipeline.}
    \label{fig:methodolgy}
\end{figure*}

In this section, we describe DP-UTIL, an extensible framework aimed for conducting comprehensive privacy/utility trade-off analysis of DP across the ML pipeline.

DP-UTIL is the first framework to combine the thus-far proposed five DP perturbation methods in a single pipeline while enabling multi-metric privacy/utility trade-off analysis. It is designed to easily add new components or update existing ones. Next, we use Figure \ref{fig:methodolgy} to describe DP-UTIL and how it can be used and extended by ML privacy practitioners or researchers. In particular, we focus on the three components: {\em DP Perturbation Plugins}, {\em Inference Attacks}, and {\em Holistic Trade-off Analysis}.

\subsection{DP Perturbation Plugins}
Across the ML pipeline, prior work has proposed five spots where DP could be plugged to enable privacy-preserving ML (hence the term ``DP Perturbation Plugins" in Figure \ref{fig:methodolgy}). Depending on dataset type (e.g., images vs. numeric), loss function (e.g., convex vs. non-convex), gradient computation method, and model architecture, the privacy guarantee offered by each DP perturbation varies. Currently, across two model architectures (LR and DNN) DP-UTIL supports five perturbation plugins: all five for LR, and input, gradient, and prediction perturbation for DNN. While our current design relies on the  peer-reviewed implementations of perturbation mechanisms for LR and DNN, users of DP-UTIL can add future implementations with minimal effort.

In terms of support for multiple datasets, currently DP-UTIL supports three datasets from vision (image classification), medical (COVID-19), and finance domains (Loan-Data) for classification tasks. The modular design allows plugging in new datasets and proceed with the rest of the analysis pipeline. The ``Pre-processing'' component in the ML pipeline block in Figure ~\ref{fig:methodolgy} is meant to offer pre-training data cleaning functionality that a user may customize depending on the dataset at hand.

\subsection{Privacy Motivated Inference Attacks}
In this component of DP-UTIL, we assume that multiple privacy-motivated inference attacks can be plugged or existing attacks can be replaced with more recent when the state-of-the-art evolves. Among inference attacks are membership inference~\cite{MIA_shokri,MIA_whitebox, MIA-Evan}, attribute inference~\cite{Property_Inference1}, model inversion~\cite{model-inversion}, and model parameter inference/extraction~\cite{model-stealing16}. In its current version, DP-UTIL supports the popular attack of membership inference attack in a black-box setting, which we introduce next. 

To uniformly analyze the utility of DP across the perturbation mechanisms, we use membership inference attack introduced by Shokri et al~\cite{MIA_shokri}. Membership inference attack aims to exploit the prediction vectors to infer the member of training datasets. For this attack, multiple shadow models are used to train an attack model. We keep our shadow models' architecture exactly the same as the target model's architecture. The attack model is a binary classifier that predicts whether a particular sample is a member of the target model's training dataset or not. We use $10$ shadow models to train our Random Forest attack model. When we attack the target models, we assume black-box access to each model, i.e., the attacker submits an input sample to a prediction API which returns prediction output.

Our choice of membership inference attack is informed by the conceptual connection to the primary goal of differential privacy, which is to make the presence/absence of a data sample indistinguishable in the eyes of an adversary. Membership inference essentially aims to achieve the opposite goal: determine, with some confidence, whether a given data sample is present or absent in a training set of a target model. This antagonistic setup between the two makes membership inference a natural fit for showcasing DP-UTIL.

\subsection{Holistic Trade-off Analysis}
Like the other components of DP-UTIL, here we envision a growing list of alternative privacy/utility trade-off analysis metrics used to evaluate model utility (e.g., via accuracy), privacy leakage, actual number of records/attributes inferred, and other relevant metrics such as performance overhead of the analysis scheme and fairness of the model predictions to a sub-population of training data (e.g., minority groups). In its current version, DP-UTIL supports three established metrics: {\em utility loss}, {\em privacy leakage}, and {\em true revealed data}, which together offer a holistic assessment of the utility of DP in limiting privacy motivated attacks such as membership inference.

\textbf{Utility Loss}. Model utility or accuracy is calculated  based on percentage rate of correctly predicted labels. We calculate utility loss or label loss as the utility difference between the non-private and the differentially private model. When utility loss is $0$, it implies that the private model achieves same utility as non-private model. Formally, utility loss is calculated as: $1 - \frac{accuracy_{private-model}}{accuracy_{non-private-model}}$.

\textbf{Privacy Leakage}. This metric~\cite{Privacy_leakage} estimates the model's susceptibility to inference attack. It quantifies the difference between true positive rate and false positive rate of the adversary's inference attack, and its value lies in the range $[0,1]$. When privacy leakage is $0$, it means that there is no data leakage induced by the inference attack, while a privacy leakage value of $1$ could essentially mean complete inference success. For some of our results, we may observe negative values for privacy leakage. In those cases, the inference attack's false positive rate is greater than true positive rates, which implies that the attack model is likely to detect more non-members as members. 

\textbf{True Revealed Data}. To quantify and observe the impact of non-members falsely inferred as members, we use the {\em true revealed data} to estimate the actual number of members whose data is in danger of disclosure when membership inference attack succeeds.

\section{Datasets and Analysis Setup}
\label{sec: dataset}
In this section, we describe the setup for our instance of DP-UTIL in Figure \ref{fig:methodolgy}. Before we describe our setup, to guide our analysis, we provide context on assumptions and scope.

\textbf{Assumptions and Scope.} We assume the correctness of the implementations of the different perturbation mechanisms we analyze. We directly use the original implementations released with the published papers. Following prior work ~\cite{DP_USENIX}, we instantiate our analysis for two classes of ML models: logistic regression (convex optimization) and deep neural networks (non-convex optimization). For logistic regression, we analyze input perturbation, objective perturbation, gradient perturbation, output perturbation, and prediction perturbation. For DNNs, we again rely on insights from prior work~\cite{DP_USENIX} that noted on the suitability of gradient perturbation for non-convex optimization techniques. In addition, we extend prior evaluations of only gradient perturbation mechanisms by introducing input perturbation and prediction perturbation as alternative DP noise mechanisms in the holistic utility analysis of DP in ML. As of this writing, we have not come across reproducible methods for objective and output perturbation for DNNS. When peer-reviewed and reproducible implementations of these missing perturbation methods are made available, the modular design of DP-UTIL allows plugging them into our holistic analysis framework to extend it to a wider range of perturbation types and their variations.

\subsection{Datasets}
We use three datasets, two of which focus on practical privacy-sensitive domains: healthcare and finance. For financial privacy analysis, we use the LendingClub-Loan dataset \cite{LendingClub} from Kaggle, while for medical privacy analysis, we use the COVID-19 dataset \cite{Covid-19}. Finally, among benchmark datasets used by prior work, we use the CIFAR-10 \cite{Cifar10} dataset. Next, we briefly describe each dataset.

\textbf{LendingClub-Loan~\cite{LendingClub}}. Lending club is a US peer-to-peer lending company that offers loans in the range $\$1,000-\$40,000$. Investors view the loan book on LendingClub website and complete their own analysis to determine the quality of the book based on the information supplied about the borrower, amount of loan, loan grade, and loan purpose. The dataset contains sensitive features about borrowers which include Zip code, employment length, loan amount, home-ownership etc. In the accepted loan data, there is a column name grade, which shows the value from A to Z,where 'A' is the highest grade and 'G' is the lowest grade. the target is to build a classifier that given the other features, classifies accepted loan into 'A' to 'G' grade. The grade is formulated using risk and volatility which adjusts final interest rates. The total number of samples in the dataset is $100K$ with $166$ features, we use $50\%$ of the dataset as training set and the remaining $50\%$ as test set.

\textbf{COVID-19~\cite{Covid-19}}. This dataset is COVID-19 related and contains sensitive information about patients as to whether a patient has underlying health conditions such as diabetics, asthma, cardiovascular, or chronic diseases. In addition, among other features, it also includes age, gender, and whether or not the patient uses tobacco. The task is a binary classification task, i.e., to predict if the patient is COVID-19 positive or negative. The dataset contains $40K$ samples with $18$ features. Similarly, we use $50\%$ of the data as training set and the remaining $50\%$ as test set.

\textbf{CIFAR-10~\cite{Cifar10}}. This dataset consists of $60K$ color images of $10$ classes. Each image has a dimension of $32 \times 32 \times 3$. The target classes include $10$ object images (e.g., airplane, bus, truck, automobile, dog, bird,frog,deer,horse, ship) that are completely mutually exclusive. We split the $60K$ samples into equal number of training and test images for our experiments. 

\subsection{Models and Hyperparameters}

\textbf{Datasets Split}. For each dataset, we first split the dataset into two: $50\%$ each. We further split the first 50\% into training and testing the model, while the remaining 50\% is also split into training and testing for the membership inference attack model. For instance, LendingClub-Loan dataset has total $100K$ samples. To train our differentially private models, we use $25K$ samples for training and $25K$ samples to test the model performance. Similarly, we use the rest of the $50K$ samples for training and testing attack models, $25K$ each. 

\textbf{Logistic Regression Model}. We train the model with $l_2$ regularization, where regularization parameters $\lambda=10^{-4}$ with $100$ epochs. For this setting, we vary our privacy budget $\epsilon$ from $10^{-2}$ to $10^{4}$. For COVID-19 datasets, $\delta=10^{-4}$ and CIFAR-10, LendingClub Loan datasets, $\delta=10^{-5}$ as it should be smaller than the inverse of each training set: $5K$, $15K$, and $25K$. Our learning rate across all datasets is $0.01$ and batch size is $250$. We use the {\em Adam} optimizer. 

\textbf{Deep Neural Network Model}. For DNN, our model has an input layer, two hidden layers, and an output layer. The input layer has the dimension of the input, the hidden layers have $64$ neurons and {\em ReLU} activation function is used. We use {\em softmax} activation function for the last layer. 

\subsection{Perturbations Setup}  
Next, we describe the specific setup we use for running the five perturbation mechanisms used in our analysis.

\textbf{Input Perturbation} To implement input perturbation for logistic regression, we use techniques from \cite{Input_perturb17}. For DNN, we implement it with respect to \cite{input-perturb20}. We use different technique for DNN as we assume DNN does not follow strong convexity considering practical cases. 
%TODO.

\textbf{Objective Perturbation}. For logistic regression, we use the Diffprivlib v0.4 library introduced by IBM \cite{Dippriv}. Their Objective perturbation technique is built based on the work of Chaudury et al. \cite{ERM-DP} and they integrate their technique with scikit-learn library under some restrictions, i.e., their logistic regression function can only perform for $l_2$ regularization.
 
\textbf{Gradient Perturbation}. For both LR and DNN, we use TensorFlow privacy framework~\cite{Tensorflow_Privacy} based on the moment accountant theory introduced by Abadi et al.~\cite{DP-SGD}. We implemented our differentially private algorithm with Gaussian Adam optimizer. To keep our privacy budget $\epsilon$ in the range $[10^{-2},10^{4}]$, we calibrated the momentum equation by only changing the noise multiplier parameter.

\textbf{Output Perturbation}. With logistic regression, we add a Gaussian noise after model parameters with the sensitivity of $\frac{2}{n\lambda}$, where $n$ is number of samples in each dataset, and we use $\lambda=10^{-4}$.

\textbf{Prediction Perturbation}. We implement PATE \cite{PATE17} proposed by Nicolas et al. More precisely, we divide LendingClub-Loan dataset, COVID-19 dataset and CIFAR10 dataset into $40$, $30$, and $40$ number of disjoint datasets, respectively, and train teacher models for each dataset. Each teacher model is trained using similar model architectures discussed earlier for both LR and DNN. To add random noise to the vote count of each label, we sample Laplace noise with privacy budget $\epsilon$ in the range $[10^{-2},10^{4}]$.

\section{Analysis Results}
\label{sec: eval}
In this section, we evaluate DP-UTIL by answering the following research questions:
\begin{itemize}
    \item \textbf{RQ1:} Among the five perturbation methods in DP-UTIL, is there a particular method that offers minimal utility loss with minimal privacy leakage?

\item \textbf{RQ2:} What is the impact of number of classes (binary vs. multi-class) on utility loss and privacy leakage across perturbation mechanisms?

\item \textbf{RQ3:} What is the impact of dataset types (image vs. numerical) on utility loss and privacy leakage across perturbation mechanisms?

\item \textbf{RQ4:} What is the impact of model architecture (shallow learning vs. deep learning) on utility loss and privacy leakage across perturbation mechanisms?
\end{itemize}

\par We now present our findings across the three datasets (CIFAR-10, COVID-19, and LendingClub-Loan), two model types (LR and DNN), and five perturbation mechanisms (input, objective, gradient, output, and prediction). We analyze utility loss (Section \ref{subsec:utility-analysis}), privacy leakage (Section \ref{subsec:privacy-leakage-analysis}), and true revealed data (Section \ref{subsec:true-positive-analysis}). 

\subsection{Utility Loss Analysis}\label{subsec:utility-analysis}

% utility loss
\begin{figure*}[t!]
\includegraphics[width=\textwidth]{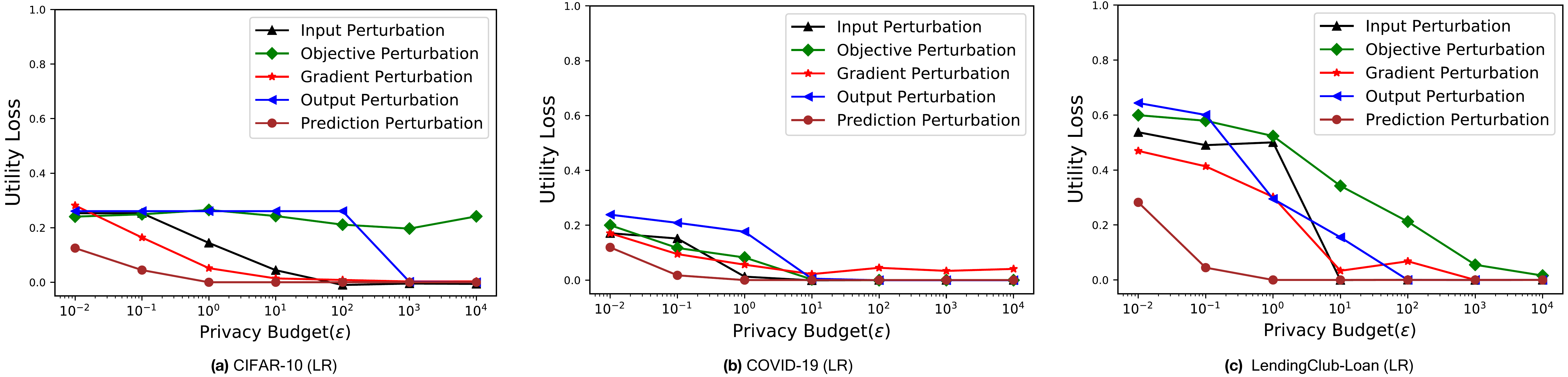}
 \vspace{.1em}
\caption{Utility loss for Logistic Regression (LR).}
\label{fig:LR-utility-loss}
\end{figure*}

\begin{figure*}[t!]
\includegraphics[width=\textwidth]{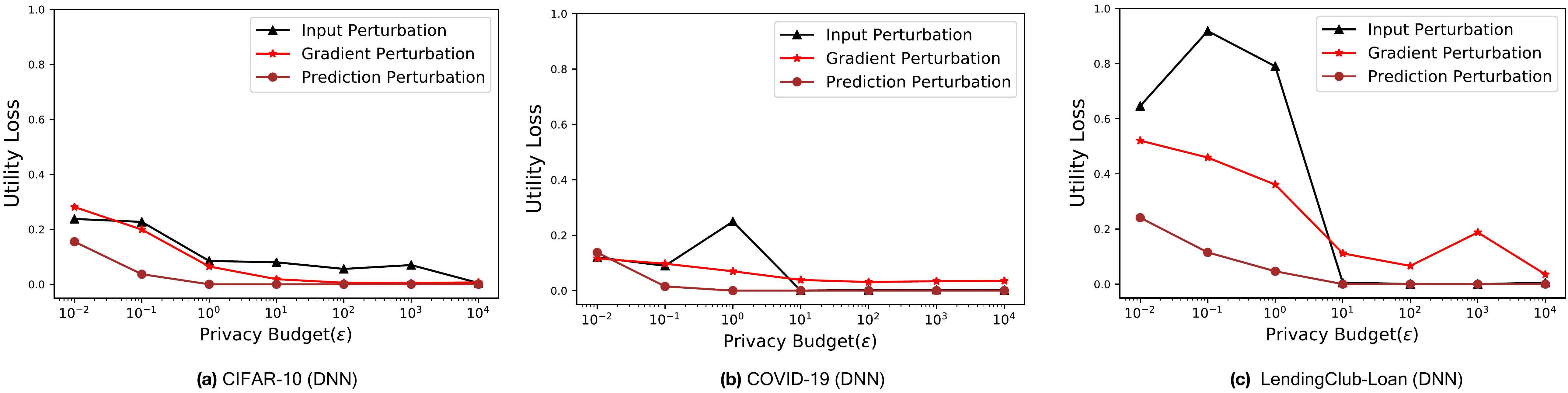}
\vspace{.1em}
\caption{Utility loss for Deep Neural Network (DNN).}
\label{fig:DNN-utility-loss}
\end{figure*}

Figures \ref{fig:LR-utility-loss} and \ref{fig:DNN-utility-loss} show utility loss ($y$-axis) for LR and DNN models, respectively, against privacy budget ($x$-axis) over $\epsilon = [10^{-2},10^{4}]$.

\textbf{Logistic Regression (LR) Utility Loss.} Figures \ref{fig:LR-utility-loss} (a), \ref{fig:LR-utility-loss} (b), and \ref{fig:LR-utility-loss} (c) show utility loss for LR  on CIFAR-10, COVID-19, and LendingClub-Loan, respectively. Among the five perturbation mechanisms, prediction perturbation consistently results in the lowest utility loss for all datasets. In fact, it incurs zero utility loss after $\epsilon = 1$ for all datasets. Next, we examine results for each dataset.

\underline{CIFAR-10}: The non-private baseline LR model achieves test accuracy of $0.375$ for CIFAR-10. From Figure \ref{fig:LR-utility-loss} (a), utility loss of output perturbation is the maximum for $\epsilon <10^3$, and this accuracy is in the range $0.01016-0.1896$. Prediction perturbation achieves utility loss $\sim0$ from $\epsilon =1$ that suggests the lowest utility loss. Objective perturbation shows utility loss of $0.25-0.23$ for $\epsilon$ in $[0,10]$ and the utility loss is $\sim0.18$ for $\epsilon>10$, which is lower compared to output perturbation. Besides, for $\epsilon > 0.1$, input and gradient perturbation show lower utility loss for small $\epsilon$ compared to objective and output perturbation. 
%Cifar10 LR:  Training:.523  Testing: 0.375  Difference:0.148

\underline{COVID-19}: The non-private LR model accuracy is $0.661$. As can be noticed from Figure \ref{fig:LR-utility-loss} (b), utility loss for output perturbation is maximum for $\epsilon <10$. For instance, at $\epsilon=1$, output perturbation has utility loss of $0.201$ which is larger compared to objective perturbation ($0.145$), gradient perturbation ($0.129$), input perturbation ($\sim0$), and prediction perturbation ($\sim0$). For $\epsilon \geq 10$, utility loss is negligible $\sim[0.04,0]$, although input, objective, output, and prediction perturbation techniques show slightly lower (almost negligible) utility loss compared to gradient perturbation. 
%Covid LR:  Training:.67  Testing: 0.66 Difference:0.01

\underline{LendingClub-Loan}: From Figure \ref{fig:LR-utility-loss} (c), we observe that for smaller $\epsilon$ value ($\epsilon<1$), output perturbation produces maximum utility loss, which is $0.665$. For all $\epsilon$ in general, we notice that prediction perturbation produces lower utility loss compared to gradient perturbation, input perturbation and objective perturbation. For example, when privacy budget $\epsilon=1$, utility loss for prediction, gradient, objective and input perturbation is $\sim0, 0.36$, $0.368$, and $0.54$, respectively. Input and gradient perturbation produce lower utility loss, $\leq 0.1$, for $\epsilon \geq 10^1$. Note that when we compare utility loss of objective perturbation with gradient and input perturbation, gradient and input perturbation are better choices than objective perturbation in terms of utility preservation. 
%Loan LR:  Training:.899  Testing: 0.8826  Difference:0.0166

\textbf{Deep Neural Network Utility Loss.} 
 Figures \ref{fig:DNN-utility-loss} (a), \ref{fig:DNN-utility-loss} (b), an \ref{fig:DNN-utility-loss} (c) show utility loss for CIFAR-10, COVID-19, and LendingClub-Loan for a DNN model on CIFAR-10, COVID-19, and LendingClub-Loan, respectively. Similar to our observation for LR, prediction perturbation incurs the lowest utility loss not only across the three datasets, but also over the whole range of privacy budget values.
 
 \underline{CIFAR-10}: The non-private model utility is $0.39$. For prediction perturbation, utility loss is $\in (0.13,0.0553)$ when $\epsilon \leq 1$, while for higher $\epsilon$ values, utility loss is $\sim0$. For gradient and input perturbation, utility loss is higher compared to prediction perturbation. For example, at lower epsilon ($\epsilon<1$), gradient and input perturbation produce $\sim0.15$ utility than prediction perturbation.
 
 %for $\epsilon < 0.01$ and $0.1$, utility loss is $0.28$ and $0.199$, respectively, while when $\epsilon \in (1,10)$, utility loss is in $(0.065,0.02)$. For higher privacy budget, $\epsilon>10$, utility loss is negligible, i.e., in the range $(0.005,0.007)$.
 
 %Cifar10 DNN:  Training:.0.521  Testing: 0.39  Difference:0.131
 
 \underline{COVID-19}: The non-private model utility is $0.6676$. We can observe from Figure \ref{fig:DNN-utility-loss} (b), utility loss in prediction perturbation is $\sim0$ for $\epsilon >0.01$. For input perturbation, utility loss is $\sim0$ for $\epsilon >1$. For gradient perturbation, utility loss is higher compared to prediction and input perturbation. For $\epsilon =0.1$ and $1$, utility loss  for gradient perturbation is $0.0972$ and $0.0698$ respectively, and for $\epsilon>1$, utility loss is $\sim0.03$. 
 
 \underline{LendingClub-Loan}: The non-private DNN model reaches accuracy of $0.71$. Comparatively, as depicted in Figure \ref{fig:DNN-utility-loss} (c), prediction and input perturbation produce lower utility loss ($\sim0$) for $\epsilon >1$ compared to gradient perturbation. On the contrary, input perturbation shows worst performance compared to other perturbation mechanisms at lower privacy budget. For $\epsilon \leq 1$, input and gradient perturbation produce $\sim0.75$ and $\sim0.4$ higher utility loss, respectively, compared to prediction perturbation.
  %For instance, for $\epsilon=1$, utility loss of prediction perturbation is $0.0466$ while utility loss of gradient perturbation is $0.36$. Prediction perturbation reaches utility loss equal $0$ at $\epsilon>1$, while for gradient perturbation, utility loss is $\sim0$ only at $\epsilon=10^4$.
 
 %Loan DNN:  Training:.7348  Testing: 0.71  Difference:0.0248

 \noindent \fbox{\parbox{0.96\columnwidth}{
 \textbf{Observation 1:}
 With regards to \textbf{RQ1}, prediction perturbation achieves the lowest utility loss across all datasets. This is intuitive as prediction perturbation requires less random noise because the noise is added to aggregated results of teachers' votes. For both LR and DNN, prediction perturbation reaches $0$ utility loss at $\epsilon \geq 0.01$ for most cases. For lower privacy budget values ($\epsilon \leq10^2$ or $\epsilon \leq10^3$), output perturbation results in the highest utility loss in contrast to gradient or objective perturbation. This result is again intuitive as objective perturbation adds noise to the objective function and afterwards minimizes the loss while output perturbation adds noise to the model parameters. Concerning \textbf{RQ2}, in LR, objective perturbation incurs more utility loss for multi-class classifiers compared to the binary-class classifier for any $\epsilon \in [10^{-2},10^4]$. Utility loss for gradient perturbation at $\epsilon <10$ shows a larger loss for multi-class classifiers compared to binary class classifiers. Additionally, input perturbation shows lower utility loss and privacy leakage for binary classifiers. In response to \textbf{RQ3}, for both LR and DNN, gradient perturbation shows lower utility loss from $\epsilon >1$ for image data compared to numerical data. Concerning \textbf{RQ4}, for gradient and prediction perturbation, we observe negligible utility loss difference between LR and DNN. Hence, for utility loss, perturbation techniques turn out to be model-independent.}}

\subsection{Privacy Leakage Analysis}\label{subsec:privacy-leakage-analysis}
% privacy leakage
\begin{figure*}[t!]
\includegraphics[width=\textwidth]{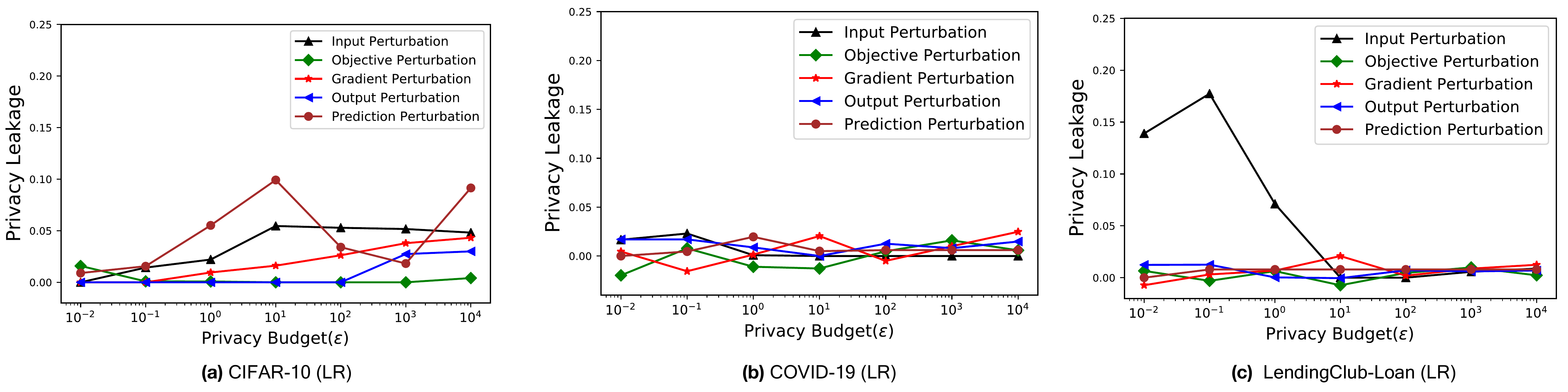}
\vspace{.1em}
\caption{Privacy leakage for Logistic Regression (LR).}
\label{fig:LR-leakage}
\end{figure*}
% \vspace{-.5em}
\begin{figure*}[t!]
\includegraphics[width=\textwidth]{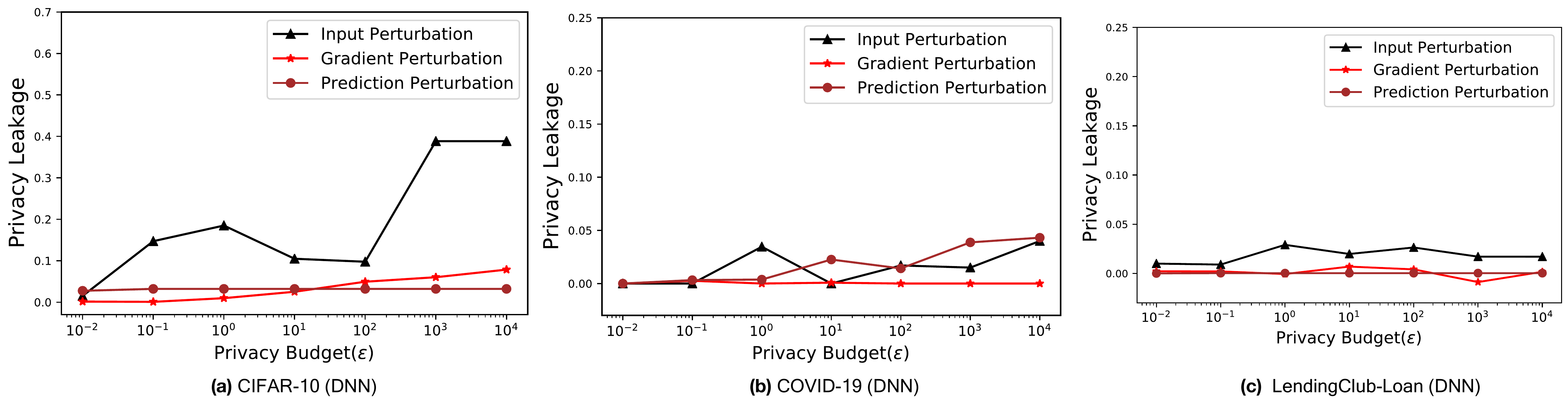}
\vspace{.1em}
\caption{Privacy leakage for Deep Neural Network (DNN).}
\label{fig:DNN-leakage}
\end{figure*}

Figures \ref{fig:LR-leakage} and \ref{fig:DNN-leakage} show the privacy leakage plots for LR and DNN models, respectively, over a range of $\epsilon$ values. 
% The $x$-axis shows privacy budget $\epsilon$ in the range $[10^{-2},10^{4}]$ and the $y$-axis shows privacy leakage of differentially private models for the perturbation mechanisms we analyse.

\textbf{Logistic Regression Privacy Leakage.} Figures \ref{fig:LR-leakage} (a), \ref{fig:LR-leakage} (b), and \ref{fig:LR-leakage} (c) show the privacy leakage  for LR model for CIFAR-10, COVID-19 and LendingClub-Loan, respectively. In the following, we analyze privacy leakage for each dataset.

\underline{CIFAR-10}: In the context of Figure \ref{fig:LR-leakage} (a), the non-private privacy leakage is $0.12$. Output perturbation shows lowest privacy leakage compared to other perturbation techniques, i.e., $0$ for $\epsilon \leq 10^2$. For $\epsilon>10^2$, privacy leakage value increases to $\sim0.03$, which is not negligible. For gradient perturbation, privacy leakage is $\sim0$ when $\epsilon <1$, while on the other hand, for higher $\epsilon$ ($\geq 1$), privacy leakage increases eventually. For example, for $\epsilon=10^4$, privacy leakage is $0.0432$ while for $\epsilon=10$, privacy leakage is $0.0162$. For objective perturbation, privacy leakage is $0.012 \pm 0.008$ over the privacy budget. Compared to the other three perturbation techniques, input and prediction perturbation show higher privacy leakage. For example, for $\epsilon \in (0,10^4)$, the privacy leakage reaches $0.054$ and $0.05$ for input and prediction perturbation, respectively.

\underline{COVID-19}: From Figure \ref{fig:LR-leakage} (b), we notice that output perturbation shows privacy leakage values $0.01\pm 0.007$ for all $\epsilon$ values. Gradient perturbation shows lower privacy leakage, $0.002 \pm 0.0026$ for $\epsilon \leq 10^2$. For larger $\epsilon$ (i.e., $ \epsilon\geq10^2$), privacy leakage reaches slightly higher estimate $0.02\pm 0.04$. For objective perturbation, privacy leakage is $\sim0$ while $\epsilon \leq 10^2$, while for larger privacy budget, leakage is $0.01 \pm 0.005$. For input perturbation, privacy leakage is $\sim0$ from $\epsilon=1$. For prediction perturbation, privacy leakage reaches $0.004\pm 0.002$ for all $\epsilon$ values. 

\underline{LendingClub-Loan}: For this dataset, input perturbation shows highest leakage, $0.1\pm 0.05$ for $\epsilon \in (0.01,1)$, otherwise $0$. Objective perturbation shows small privacy leakage over all $\epsilon$ values, which is $0.006\pm 0.003$. For gradient perturbation, privacy leakage value is $0.003 \pm 0.003$ for $\epsilon \leq 1$. For $\epsilon \geq 1$, this value reaches $0.014 \pm 0.007$. For output perturbation, privacy leakage does not follow a pattern for all $\epsilon$ and it varies from $0.007 \pm 0.0052$. For prediction perturbation technique, privacy leakage is $0$ for $\epsilon=0.01$, and $0.00792$ for $\epsilon \in (0.1,10^4)$.

\textbf{Deep Neural Network Privacy Leakage.}
Figures \ref{fig:DNN-leakage} (a), \ref{fig:DNN-leakage} (b), and \ref{fig:DNN-leakage} (c) show the privacy leakage of a DNN model for CIFAR-10, COVID-19 and LendingClub-Loan, respectively. 

\underline{CIFAR-10}: As can be seen from Figure \ref{fig:DNN-leakage} (a), input perturbation shows higher leakage compared to gradient and prediction perturbation. Privacy leakage is $\in(0.01,0.15)$ for $\epsilon <10^2$ and $\sim0.38$ for higher privacy budget values. Gradient perturbation shows comparatively lower privacy leakage $<0.01$ for $\epsilon\leq10$. For privacy budget value higher that that, privacy leakage is incremental. For example, at $\epsilon=10$, privacy leakage is $0.0251$ which reaches $0.0785$ at $\epsilon=10^4$. Also note that, for gradient perturbation, privacy leakage value drops compared to prediction perturbation at $\epsilon \leq 10$, where as from $\epsilon > 10$, privacy leakage is slightly lower for prediction perturbation. At $\epsilon=1$ privacy leakage is $0.0097$ for gradient perturbation while for prediction perturbation this value reaches $0.032$. But for higher $\epsilon$ (i.e., $\epsilon=10^3$), privacy leakage is $0.0602$ for gradient perturbation while for prediction perturbation this value reaches $0.0321$.

\underline{COVID-19}: From Figure \ref{fig:DNN-leakage} (b), prediction perturbation results in more privacy leakage compared to gradient and input perturbation ($\sim0$ and $\in (.03\pm .02)$, respectively). For prediction perturbation, privacy leakage is incremental with respect to increasing value of $\epsilon$. For instance, at $\epsilon=0.1$, privacy leakage is $0.003239$ while for $\epsilon=10^4$, privacy leakage reaches $0.0431$.
%Training:0.6942  Testing: 0.67  Difference:0.034

\underline{LendingClub-Loan}: As can be seen from Figure \ref{fig:DNN-leakage} (c), gradient perturbation shows privacy leakage $\sim0$ over all the $\epsilon$ values. For instance, for $\epsilon=10^4$, privacy leakage is $0.00164$. For prediction perturbation, privacy leakage is also $\sim0$ though slightly larger for several $\epsilon$ values. For $\epsilon=10^3$, privacy leakage of prediction perturbation is $\sim0.001$ while gradient perturbation reaches $\sim0$. Input perturbation shows highest privacy leakage, which is $\sim0.04$ for almost all values of $\epsilon$.

 \fbox{\parbox{.96\columnwidth}{
 \textbf{Observation 2:} In response to \textbf{RQ1}, for LR, objective perturbation shows the lowest privacy leakage compared to other perturbation techniques, which is no leakage for almost all $\epsilon$ choices. On the contrary, for DNN models, gradient perturbation is the best for a privacy practitioner while considering privacy leakage, as leakage seems negligible for different $\epsilon$ choices. Concerning \textbf{RQ2}, LR for the binary classifier shows almost $0$ privacy leakage at $\epsilon<10^3$, which seems promising as it shows almost no privacy leakage with lower utility loss in contrast to multi-class classifiers. In response to \textbf{RQ4}, input perturbation for DNN shows more leakage in contrast to LR for higher privacy budget values (i.e., $\epsilon\sim10$), which is expected since we use two different input perturbation mechanisms (input perturbation for DNN follows more relaxed boundaries).}}

\subsection{True Revealed Records}\label{subsec:true-positive-analysis}

% true positive
\begin{figure*}[t!]
\includegraphics[width=\textwidth]{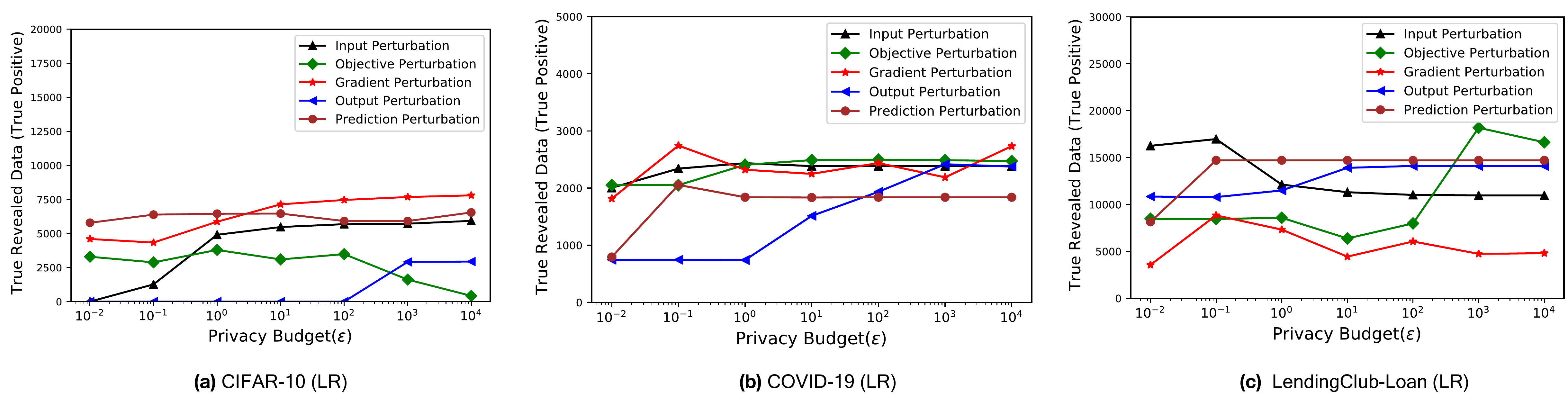}
\vspace{.1em}
\caption{True revealed data for Logistic Regression (LR).}
\label{fig:LR-tp}
\end{figure*}

\begin{figure*}[t!]
\includegraphics[width=\textwidth]{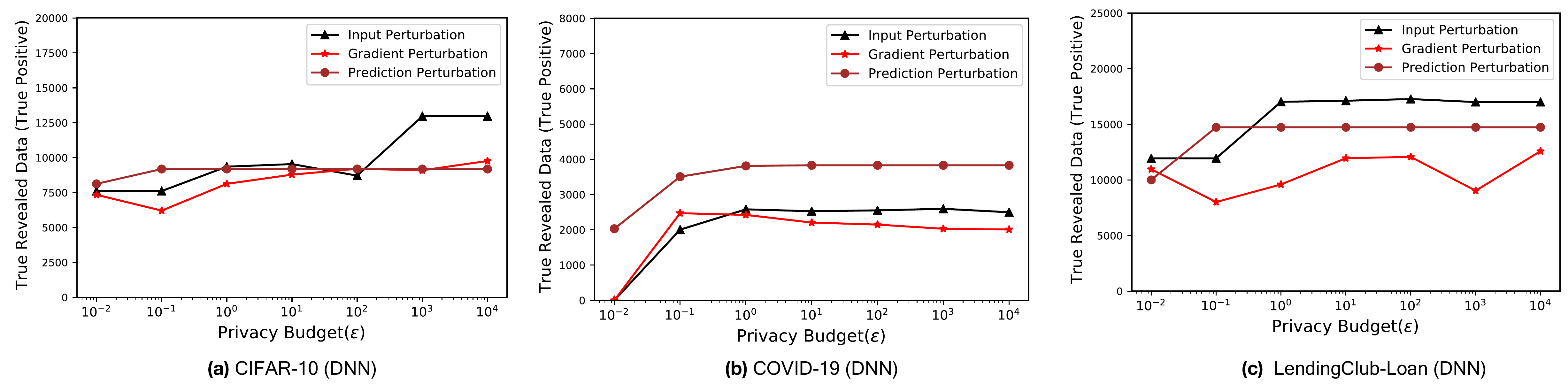}
\vspace{.1em}
\caption{True revealed data for Deep Neural Network (DNN).}
\label{fig:DNN-tp}
\end{figure*}

Figures \ref{fig:LR-tp} and \ref{fig:DNN-tp} show true revealed record/true positive value with respect to privacy budget $\epsilon$ for CIFAR-10, COVID-19, and LendingClub-Loan, respectively. 

\textbf{Logistic Regression True Revealed Records}. 
Figures \ref{fig:LR-tp} (a), \ref{fig:LR-tp} (b), and \ref{fig:LR-tp} (c) show the true revealed records of a LR model over CIFAR-10, COVID-10, and LendingClub-Loan, respectively. Next, we examine results for each dataset. 

\underline{CIFAR-10}: From Figure \ref{fig:LR-tp} (a), prediction perturbation shows highest true positive data leakage ($10368$) for overall privacy budget metrics. Output perturbation shows lowest revealed true positive records, surprisingly ($0$), for $\epsilon <10^3$. For privacy budget $\epsilon \geq 10^3$, this value is $\sim3,000$. For gradient perturbation, this value reaches $7,000-8,000$ after $\epsilon>1$. Objective and input perturbation also reveal lower number of true positive values compared to gradient and prediction perturbation which is between $(\sim6,000 \pm 1000)$ over the privacy budget range. 

\underline{COVID-19}: According to Figure \ref{fig:LR-tp} (b), while total number of training data is $5000$ and $\epsilon <10$, output perturbation shows lowest values $(\leq 750)$ compared to other perturbation techniques, hence for $\epsilon \geq 10$, true positive value increases eventually. For instance, for $\epsilon=100$, true revealed value is $\sim2,000$ which is $40\%$ of total number of training data. On the other hand, for prediction perturbation, true revealed value is $\sim1850$ when $\epsilon > 0.01$. Objective, gradient, and input perturbation show close numbers of revealed members which is higher than prediction perturbation. For example, gradient perturbation reveals $\sim2,5000$ true positive values from $\epsilon>0.01$.
%true revealed samples are $\sim2500$ for $\epsilon \geq 1$, while for $\epsilon < 1$, true number of true positive value is $\sim2,000$. 

\underline{LendingClub-Loan}: This result is shown in Figure \ref{fig:DNN-tp} (c).  When total number of training data is $25000$, for $\epsilon \in (0.01,10^4)$, for output perturbation this value increases from $10848$ to $14101$. For instance, for $\epsilon=100$, true revealed value is $14113$ which is $56\%$ of total number of training data. On the other hand, for prediction perturbation, it is $14715$ when $\epsilon > .01$, which is the highest. In this context, input perturbation shows better performance compared to output and prediction perturbation. For objective perturbation, true positive samples or true revealed samples are $\sim8,000$ for $\epsilon \leq 100$, while for $\epsilon > 100$, total number of true revealed value is $\sim17,000$. For gradient perturbation, true positive value is $\sim6000 \pm 2000$ from $\epsilon>0.01$, which is the lowest and almost constant over different $\epsilon$.

\textbf{Deep Neural Network True Revealed Records.}
 Figures \ref{fig:DNN-tp} (a), \ref{fig:DNN-tp} (b), and \ref{fig:DNN-tp} (c) show revealed true members of training dataset or true positive values for a DNN model over CIFAR-10, COVID-19, and LendingClub-Loans, respectively.
 
 \underline{CIFAR-10}: From Figure \ref{fig:DNN-tp} (a), gradient perturbation reveals lower true positive value than prediction and input perturbation for $\epsilon \leq 10^2$, while for $\epsilon >10^2$, the numbers are nearly equal for both gradient and prediction perturbation, while input perturbation reveals larger values. For example, at $\epsilon=1$, gradient perturbation leaks $8126$ true positive values while prediction perturbation and input perturbation reveals $\sim9200$ and $\sim9300$ values respectively. 
 
\underline{COVID-19}: We observe from Figure \ref{fig:DNN-tp} (b) that prediction perturbation revealed more true positive values compared to gradient perturbation and input perturbation for all $\epsilon$. For example, at $\epsilon=100$, true revealed value is $3831$ for prediction perturbation while gradient and input perturbation reaches $2148$ and $2552$ respectively.

\underline{LendingClub-Loan}: Figure \ref{fig:DNN-tp} (c) shows that true positive value of input and prediction perturbation is higher than gradient perturbation. For example, at $\epsilon=10^3$, prediction and input perturbation reaches true positive value of $14,733$ and $17000$ while gradient perturbation reaches $9034$. 

\noindent \fbox{\parbox{.96\columnwidth}{
 \textbf{Observation 3:} True revealed records has almost a linear relationship with privacy leakage. Over all the results, we observe that a model starts to leak more true records when the privacy leakage is higher.}} 
 
 \subsection{Overall Observations on Utility/Privacy Trade-offs}
  With regards to \textbf{RQ1}, if we contemplate overall performance (considering both utility and privacy), there is no obvious optimal DP technique that fits well for LR. On the other hand, for deep learning models, gradient perturbation seems an obvious choice a practical utility/privacy trade-off. It is also noticeable that, for gradient perturbation, privacy budget $\in [1,10^2]$ provides acceptable privacy utility trade-offs. We note that our results so far do not point to a reality where one perturbation technique offers better/acceptable utility at no cost (compromise on privacy is inevitable). For instance, prediction perturbation provides better utility compared to other perturbation techniques, but it costs the highest privacy leakage in exchange. 
  
  Concerning \textbf{RQ2} and analyzing Observation 1 and 2, for a privacy practitioner who wants to work with a binary classifier and LR model, objective perturbation is an optimal choice. However, other perturbation techniques, for instance, gradient perturbation, are better choices for multi-class classifiers compared to objective perturbation as the utility/privacy trade-off is within a tolerable range for gradient perturbation. 
  
  In response to  \textbf{RQ3}, analyzing observation 1 and 2, we conclude the overall better performance of gradient perturbation on image datasets compared to numerical datasets. 
  
  Concerning \textbf{RQ4}, we do not observe fluctuations for different perturbation techniques considering different model architectures.

\section{Related Work}
\label{sec: related}
 While previous studies evaluated privacy-accuracy trade-off in terms of privacy budget for different perturbation techniques, they do them in isolation, for example, performing studies for only gradient perturbation. In this context, the absence of a comprehensive picture of privacy/accuracy trade-off for widely adopted perturbation techniques over the ML pipeline is what makes our work broadly orthogonal to prior work ~\cite{DP_USENIX,input-perturb20,DPUtility20}. In the following, we highlight the most relevant related works. 

Early usages of differential privacy for privacy-preserving ML include empirical risk minimization \cite{ERM-DP,input-perturb20,Faster_ERM} and designing differentially private deep learning algorithms \cite{DP-SGD,Denoising-Nsr}. These class of differentially private algorithms add noise in different stages of the ML pipeline via input perturbation~\cite{Input_perturb17,input-perturb20,Input-Perturb13}, objective perturbation~\cite{ERM-DP}, gradient perturbation~\cite{DP-SGD}, output perturbation~\cite{ERM-DP}, and prediction perturbation~\cite{PATE_new,PATE17}, confidence masking~\cite{Memguard}.

In \cite{Input_perturb17}, a DP technique for input perturbation is proposed. In this work, they inject random noise into the input data in a manner that satisfies $(\alpha\epsilon, \delta)$ local differential privacy to the database and $(\epsilon,\delta)$ global differential privacy to the model parameters. In \cite{input-perturb20}, they expanded the previous works limited to strong convex loss function using the Polyak-Lojasiewicz \cite{Polyak-Lojasiewicz} condition.
%in which they show that their input perturbation technique perturbs the gradient and model parameters. 

Objective perturbation for Empirical Risk Minimization (ERM) was proposed by Chaudhury et al. \cite{ERM-DP}. This work assumes several convexity and differentiability criteria, i.e., strictly convex function and normalized input data. They expand their technique to produce privacy-preserving LR and Support Vector Machine. In \cite{Faster_ERM}, a differentially private ERM is studied for strongly convex loss function with or without non-smooth regularization. Their contribution is that they improve gradient complexity as $O(n\log{n})$. Besides, for high dimensional data, they reduced the gradient complexity to $O(n^2)$, which is more general and faster than previous works. In a recent work, \cite{Distributed_learning}, gradient perturbation on collaborative learning is introduced where multiple data owners add noise to their respective gradient locally after each iteration. 

In \cite{Memguard}, they add a carefully crafted noise vector obtained from a mechanism using an adversarial example to defend membership inference attack. The goal is to choose random noise that minimizes membership inference attack accuracy while keeping the true label. In \cite{PATE17}, \cite{PATE_new},  an ensemble of teacher models is trained using disjoint datasets. A separate model, called the student model, is trained-based data labeled with output obtained from a noisy aggregation of the prediction results of teacher ensembles. The main difference between PATE \cite{PATE17} and Scalable PATE \cite{PATE_new} is the latter uses more concentrated noise (Gaussian noise) while the former uses Laplace noise. Besides, the latter is more selective (i.e., in case of more disagreements among the teachers, the system may simply choose to abstain). In PRICURE ~\cite{PRICURE21}, a similar strategy to PATE is used to add noise to an aggregation of predictions from multiple models in a collaborative setting so as to limit the success of membership inference attack.

\par Jayaraman and Evans \cite{DP_USENIX} evaluate relaxed notions of DP mechanisms for ML. Using gradient perturbation as the DP mechanism, they focus on three relaxations: DP with advanced composition~\cite{Advanced-Comp},
zero-concentrated DP~\cite{Zero-Conc-DP}, and Rényi DP~\cite{Renyi-DP}. This work explores utility/privacy trade-off through leakage measurement due to these relaxed notions of DP. Their results conclude that existing DP-ML methods rarely offer acceptable privacy-utility trade-offs for complex models. 

Another evaluation of DP over healthcare dataset was studied by Vinith et al. \cite{DP_Healthcare}. They studied DP-SGD models in clinical prediction tasks such as  X-ray classification and mortality prediction. Their work concludes that DP-SGD loses salient information about minority classes while it preserves data privacy. In \cite{Unintended_Memorization}, they evaluate the effect of differential privacy memorization attack, though does not provide privacy leakage evaluation. In \cite{MIA_experiment}, they perform measurement studies on the upshot of DP against MIA and this evaluation was limited to DP-SGD \cite{DP-SGD}. 

In \cite{DPUtility20}, a similar analysis pipeline as ours is presented. They choose Naive Bayes for input and output perturbation and NN for input and gradient perturbations to evaluate utility/privacy trade-offs on CIFAR, Purchase, and Netflix datasets. The main difference of this work with ours is we evaluated five perturbations for LR and the currently popular three perturbations for DNN, hence DP-UTIL enables holistic analysis.

\section{Conclusion}
\label{sec: concl}
 In this paper, we introduced a holistic utility analysis of differential privacy over the machine learning pipeline. Our principal contribution is the holistic analysis of five existing DP-perturbations on logistic regression and three existing DP perturbations on deep neural networks through a comprehensive privacy/utility trade-off analysis over a range of privacy budget $\epsilon$. 
From our evaluations, we observe that some perturbation mechanisms outperform others in terms of utility, yet cost privacy leakage in exchange. Besides, our results also offer insights into how DP-techniques compare across different datasets and classifiers. We also report negligible differences in terms of utility and privacy over diverse model architectures. For example, for deep neural networks, gradient perturbation offers an acceptable utility/privacy trade-off over other perturbation methods, where as for binary classifiers and LR, objective perturbation provides acceptable utility-privacy compared to multi-class classifiers. 
We hope our holistic analysis framework will enable machine learning privacy practitioners to make informed decisions as to which perturbation mechanism to pick based on thorough comparative analysis of the dynamics between optimization techniques in machine learning, perturbation mechanisms, number of classes, and privacy budget.  
%In this paper, we introduced a holistic utility analysis of differential privacy over the machine learning pipeline. Using utility loss, privacy leakage, and the number of truly revealed data samples, we analyze input perturbation, objective perturbation, gradient perturbation, output perturbation, and prediction perturbation over a range of $\epsilon$ values. From our evaluations, we can advocate that assorted perturbation mechanisms outperform others in terms of utility, yet cost privacy leakage in exchange. Besides, our results interpret a comparative analysis and show how DP-techniques fluctuate over different model architecture, data, and classifiers. We hope our holistic analysis framework will enable machine learning privacy practitioners to make informed decisions as to which perturbation mechanism to pick based on thorough comparative analysis of the dynamics between optimization techniques in machine learning, perturbation mechanisms, number of classes, and privacy budget. 

%submit what we have, Input perturbation /objective perturbation, output perturbation to a journal Dataset Devs paper, Add attribute inference attack
% For input perturbation, for image datasets (e.g., CIFAR10), we can try performing minimal perturbations in a label-preserving manner on each input see the impact and compare it against the other type of perturbations
% For the current results, we probably need k number of runs for each experiments to see if the results are fairly stable. In the USENIX'19 paper, they do run each experiment for k =5.
\bibliographystyle{ACM-Reference-Format}
\bibliography{main}
\end{document}